%% file: main.tex
\documentclass{article}
\usepackage{spconf,amsmath,graphicx}
\usepackage[absolute]{textpos}

\usepackage{xcolor}
\usepackage{subcaption}
\usepackage{hyperref}
\usepackage{graphicx}
\usepackage{amsmath}
\usepackage{stackengine} 
\usepackage{MnSymbol}
\usepackage{multirow}

\newcommand \ignore[1]{}

\usepackage{fancyhdr}
 \fancypagestyle{firststyle}
 {
 \fancyhf{} % sets both header and footer to nothing
 \fancyfoot[LO]{© 2023 IEEE. Personal use of this material is permitted. Permission from IEEE must be obtained for all other uses, in any current or future media, including reprinting/republishing this material for advertising or promotional purposes, creating new collective works, for resale or redistribution to servers or lists, or reuse of any copyrighted component of this work in other works. }
 }
\title{mLCANets: Multi-layer Lateral Competition Neural Network-based robust audio classification}

\title{LCANets++:  Robust audio classification using Multi-layer Neural Networks with Lateral Competition}
\name{Sayanton V. Dibbo$^{\mathsection \dagger}$\sthanks{Author performed the work
	while working at the Los Alamos National Laboratory}, Juston S. Moore$^{\mathsection}$, Garrett T. Kenyon$^{\mathsection}$,  Michael A. Teti$^{\mathsection}$} \address{$^{\mathsection}$ Los Alamos National Laboratory, Los Alamos, NM, USA, $^{\dagger}$Dartmouth College, Hanover, NH, USA}
\ignore{+++++++++++++++++

\twoauthors
  {Sayanton V. Dibbo, Juston S. Moore, Garrett T. Kenyon}
	{School A-B\\
	Department A-B\\
	Address A-B}
  {Michael A. Teti\sthanks{The fourth author performed the work
	while at ...}}
	{School C-D\\
	Department C-D\\
	Address C-D}
 +++++++++++++++}

\begin{document}
%\ninept
%

\maketitle

 \thispagestyle{firststyle}

\begin{abstract}
%%%%%%%%  Version 5  %%%%%%%%
%%%%%  148/150 words limit   %%%%%%%ICASSP 2024%
Audio classification aims at recognizing audio signals, including speech commands or sound events. However, current audio classifiers are susceptible to perturbations and adversarial attacks. In addition, real-world audio classification tasks often suffer from limited labeled data. To help bridge these gaps, previous work developed neuro-inspired convolutional neural networks (CNNs) with sparse coding via the Locally Competitive Algorithm (LCA) in the first layer (i.e., LCANets) for computer vision. LCANets learn in a combination of supervised and unsupervised learning, reducing dependency on labeled samples. Motivated by the fact that auditory cortex is also sparse, we extend LCANets to audio recognition tasks and introduce LCANets++, which are CNNs that perform sparse coding in multiple layers via LCA. We demonstrate that LCANets++ are more robust than standard CNNs and LCANets against perturbations, e.g., \textit{background noise}, as well as black-box and white-box attacks, e.g., \textit{evasion} and \textit{fast gradient sign (FGSM)} attacks.
\end{abstract}
\input{Introduction}

\input{Methods}
\input{Experiment}

\input{Results}
\input{Conclusion}
\input{Sections/6.Acknowledgements}

\newpage
\bibliographystyle{IEEEbib}
\bibliography{refs}

\end{document}

%% file: Introduction.tex
\section{Introduction}
\label{intro}
Audio signal classification for the purposes of sound recognition (SR) or sound event detection (SE) has become an active area of research interest~\cite{solovyev2020deep, hershey2017cnn}. This includes using the Convolutional Neural Network (CNN) models for understanding human speech words, e.g., `yes', `stop', etc., or classifying sound events like `baby cries', and `barking'. However, standard CNNs are notoriously susceptible to perturbations or adversarial attacks~\cite{carlini2018audio,  xie2021enabling, dibbo2023model, dibbo2023sok}. Standard audio classification models depend highly on large labeled datasets for better performances~\cite{wang2021few, dibbo2022phone}, but large labeled datasets can be scarce for many common tasks, such as speaker identification. Generating augmented samples for audio data is one proposed approach to mitigate this challenge, but data augmentation can be time-consuming and expensive~\cite{koyama2022spatial, bautista2022speech}. Therefore, it is crucial to develop audio classifiers that can learn robust features with limited labeled samples.  

% Researchers have found that although the CNN models achieve good performances, they are susceptible to perturbations and adversarial attacks~\cite{szegedy2014intriguing, goodfellow2014explaining}. In particular, for CV tasks, CNN models exhibit different behavior than biological vision systems and are less robust against perturbations or attacks~\cite{teti2022lcanets}. However, robustness of CNN  models in audio classification tasks is still unclear. 

Recent studies have shown that CNNs that are more similar to the primary visual cortex are more robust than standard CNNs \cite{dapello2020simulating}. Based on this, previous work developed CNNs in which the first layer performed sparse coding via the Locally Competitive Algorithm (LCA) \cite{paiton2020selectivity,teti2022lcanets, li2022revisiting}, which is a biologically plausible model of the primate primary visual cortex \cite{olshausen1996emergence}. These CNNs, which we refer to as LCANets, were shown to be more robust than standard CNNs on standard CV tasks. However, there are two issues with this approach we address here. First, sparse coding models were designed to model the visual cortex, so it is unclear how they will impact the performance of CNNs on audio classification tasks. Second, these LCANets were robust to natural corruptions, but they were susceptible to white-box adversarial attacks \cite{teti2022lcanets} unless the exact attack was known before hand \cite{li2022revisiting}.

% This uniqueness of sparse coding makes researchers analyze the robustness of these models by incorporating sparse coding via the Locally Competitive Algorithm (LCA) in the first layer of the CNN models in CV~\cite{teti2022lcanets}. These LCANets show promising robustness in CV. Being inspired by this direction, for the first time in this paper, we introduce sparse coding via LCA in CNN models (LCANets) for the audio classification tasks.

\begin{figure*}
%\centering
\begin{subfigure}{.5\textwidth}
  \centering
  \includegraphics[width=8cm, height=2.5cm]{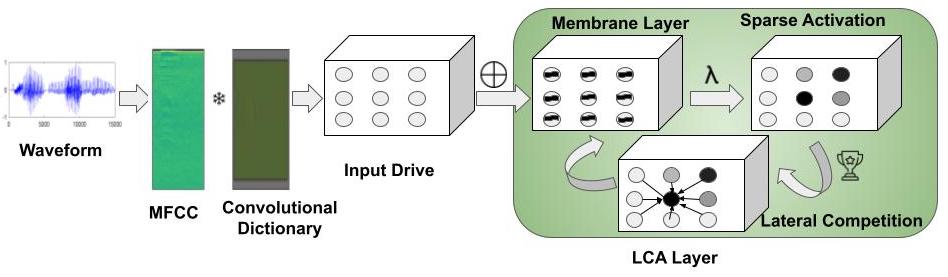}
  \caption{LCA Frontend.}
  \label{fig:lca_frontend}
\end{subfigure}%
\begin{subfigure}{.5\textwidth}
  \centering
  \includegraphics[width=1.0\linewidth, height=3.5cm]{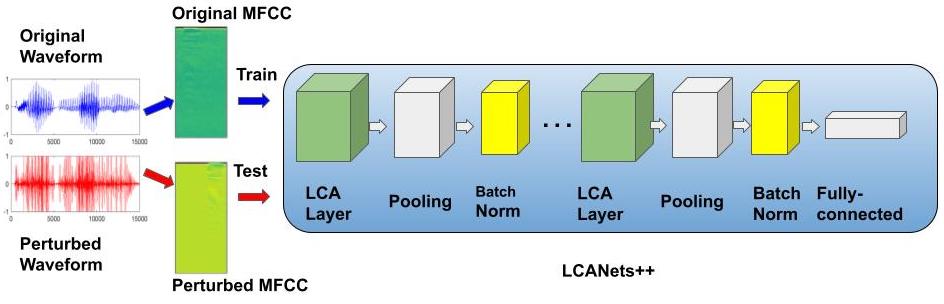}
  \caption{The overall architecture of our proposed LCANets++.}
  \label{fig:lca_pp}
\end{subfigure}
\caption{An overview of (a.) LCA frontend and (b.) pipeline of our proposed LCANets++, utilizing sparse coding via multiple LCA layers in the state-of-the-art (SOTA) CNN backbone, enabling lower misclassification on perturbed test sets or attacks.}
\label{fig:overview_lcanets++}
\end{figure*}

% In CV, models with LCANets are robust against noise and perturbations. Nevertheless, these models do not demonstrate robustness against white-box adversarial attacks~\cite{teti2022lcanets}. 

Motivated by this, we introduce multi-layer LCANets, which perform sparse coding in multiple CNN layers. We refer to these multilayer LCANets as LCANets++ and train them on audio classification tasks. To test the robustness of LCANets++ relative to LCANets and standard CNNs, we  first conducted experiments with different audio perturbations, e.g., \textit{background noise}. In addition, we show that our proposed LCANets++ are more robust compared to the state-of-the-art (SOTA) models (e.g., ResNet18, standard CNN) and LCANets against white-box attacks, i.e., \textit{fast gradient sign attack (FGSM)}~\cite{kurakin2018adversarial} and \textit{projected gradient descent attack (PGD)}~\cite{chiang2020witchcraft}, as well as black-box attacks, i.e., \textit{evasion attack}.

%% file: Methods.tex
\section{Proposed Method}
\label{sec: method}
% In this section, LCA frontend, LCANets, and our proposed LCANets++ for audio classification tasks will be introduced. 

\subsection{LCA Layer}\label{subsection:lca_layer}
As presented in Fig.~\ref{fig:overview_lcanets++}, LCA layer is the basic building block for the LCA frontend and our proposed LCANets++. LCA layer converts the input $\mathcal X$ to a coding $\mathcal C$, i.e., representation of the input $\mathcal X$, leveraging the least number of active neurons (i.e., features). The goal of the reconstruction minimization problem applied here is to find the sparse coding representation $C$ closest possible to original input $\mathcal X$ as follows:
\begin{equation}\label{eqn:re_loss}
\mathcal L_{re}=\min_{C}\frac{1}{2} || \mathcal X- \mathcal C \ostar \Phi ||_2^2 + \lambda || \mathcal C ||_1
\end{equation}
where $\mathcal L_{re}$ denotes the reconstruction loss, $\mathcal X$ is the original input, $\mathcal C$ is the sparse code, $\ostar$ is transpose convolution, $\Phi$ is dictionary components learned last iteration, and $\lambda$ is the trade-off (i.e., regularization) constant. LCA layers perform lateral competitions to fire neurons and a neuron membrane follows the following ordinary differential equation~\cite{teti2022lcanets}:
\begin{equation}\label{eqn:ode_membrane}
\hat {\mathcal M(t)}=\frac{1}{\gamma} [\mathcal D(t)-\mathcal M(t)- \mathcal C(t)*\mathcal S + \mathcal C(t)]
\end{equation}
where $\gamma$ is time constant, $\mathcal D(t)$ stands for neuron's input drive obtained by convolution of inputs with  dictionary, i.e., $\mathcal X \star \Phi$; $\mathcal M(t)$ is neuron's membrane potential, $\mathcal S$= $\Phi \star \Phi$ is the pairwise feature similarity, and $\mathcal C(t)$ is the neuron's firing rate obtained by applying a soft threshold activation on the membrane potential $\mathcal M(t)$. Coordinate ascent is used to learn dictionary $\Phi$, which solves for $\mathcal C$, given an input batch using LCA and then updates $\Phi$ with stochastic gradient descent (SGD).
%\end{document}

\subsection{LCA Frontend}\label{subsection:lca_frontend}
LCA frontend is the unsupervised pre-training part of the LCANet, as shown in Fig.~\ref{fig:lca_frontend}. It basically consists of the raw audio waveform converted to MFCCs, as input signal $\mathcal X$ and the LCA layer to compute the sparse representation $\mathcal C$ of the input $\mathcal X$, which can then feed to conventional CNN layers for the classification task.  

\subsection{LCANets}\label{subsection:lca_nets}
LCANets for the audio classification consist of the LCA frontend and followed by CNN layers. LCA frontend learns in unsupervised fashion and then passes the computed sparse code $\mathcal C$ to the CNN layers to finally perform the classification task. One major difference is that, the sparse code $\mathcal C$ does not need to recompute back to original input $\mathcal X$ before feeding to CNN layer, as other reconstruction-based models usually do. This makes the LCANets more effective agaisnt perturbations, while reducing dependency on labeled audio samples.

\subsection{LCANets++}\label{subsection:lca_pp}

We present the overview of our proposed LCANets++ in Fig.\ref{fig:lca_pp}. The basic building block of our proposed LCANets++ is the LCA layers. In this architecture, multiple LCA layers are inserted that learn in unsupervised fashion. The convolutional layers in SOTA CNN networks are replaced by the LCA layers, performing sparse coding in each layer. Similar to~\cite{chen2022minimalistic}, in order to reduce over-sparsity, in between two consecutive sparse layers (i.e., LCA layers), a dense layer, i.e., batch normalization layer, is mounted (Fig.\ref{fig:lca_pp}) in our LCANets++.  

%% file: Experiment.tex
\section{Experiments}
\label{exp}
In this section, details of the experimental setup, including dataset, pre-processing, and models, are described.

\subsection{Dataset and Pre-processing}
\label{subsec:exp_dataset}
We experiment with Google Speech Commands v2~\cite{warden2018speech} dataset. This dataset has audio waveforms of 35 classes of human speech commands like ``yes," ``no," and ``left", ``right." We perform pre-processing on the raw waveforms of the three influential classes, i.e., ``yes," ``no," and ``stop" to obtain the Mel-frequency cepstral coefficient (MFCC)~\cite{dibbo2022phone} features and train all the models with the MFCCs of the waveforms.

\subsection{Models}
\label{subsec:exp_models} We experiment with regular CNN models with 2 convolutional layers. In our LCANets++ on CNN model, we replace both convolutional layers with the LCA layers. We also compare our LCANets++ with the larger SOTA model, i.e., ResNet18. In the ResNet18 model, we replace the alternative convolutional layers in the first block with LCA layers to obtain the ResNet18\_LCA++ model. To experiment with the performance of LCANets++ against \textit{white-box} or \textit{black-box} attacks, we consider the regularization constant $\lambda=1.00$ for better sparse representations and hence, improved robustness against perturbations or attacks.

\subsection{Experimental Setup}
\label{subsec:exp_setup} We run all the experiments on 8 nodes NVIDIA\_A100-PCIE-40GB GPUs with 64-128 cores on the cluster. We use the Pytorch framework to develop the LCA class and LCANets++ implementations. We consider a train test split of 70\% and 30\% for all the models experimented in this work. For the \textit{background noise} experiment, we train models for 50 epochs, for rest of the experiments models are trained with 20 epochs.  We consider 0.0001 learning rate. We use SGD optimizer with 0.9 momentum for optimization. In order to add background noise, we impose background noise on all test set raw waveforms of audio clips, tuning the SNR [db] values to obtain different perturbed test sets. Similarly, we consider perturbing MFCCs with different $\epsilon$ values.

%% file: Results.tex
\section{Results and Analysis}
\label{result}
In this section, we illustrate the key results of our experiments on different perturbations and adversarial attacks.

\begin{figure}
%\centering
\begin{subfigure}{.24\textwidth}
  \centering
  \includegraphics[width=4.0cm]{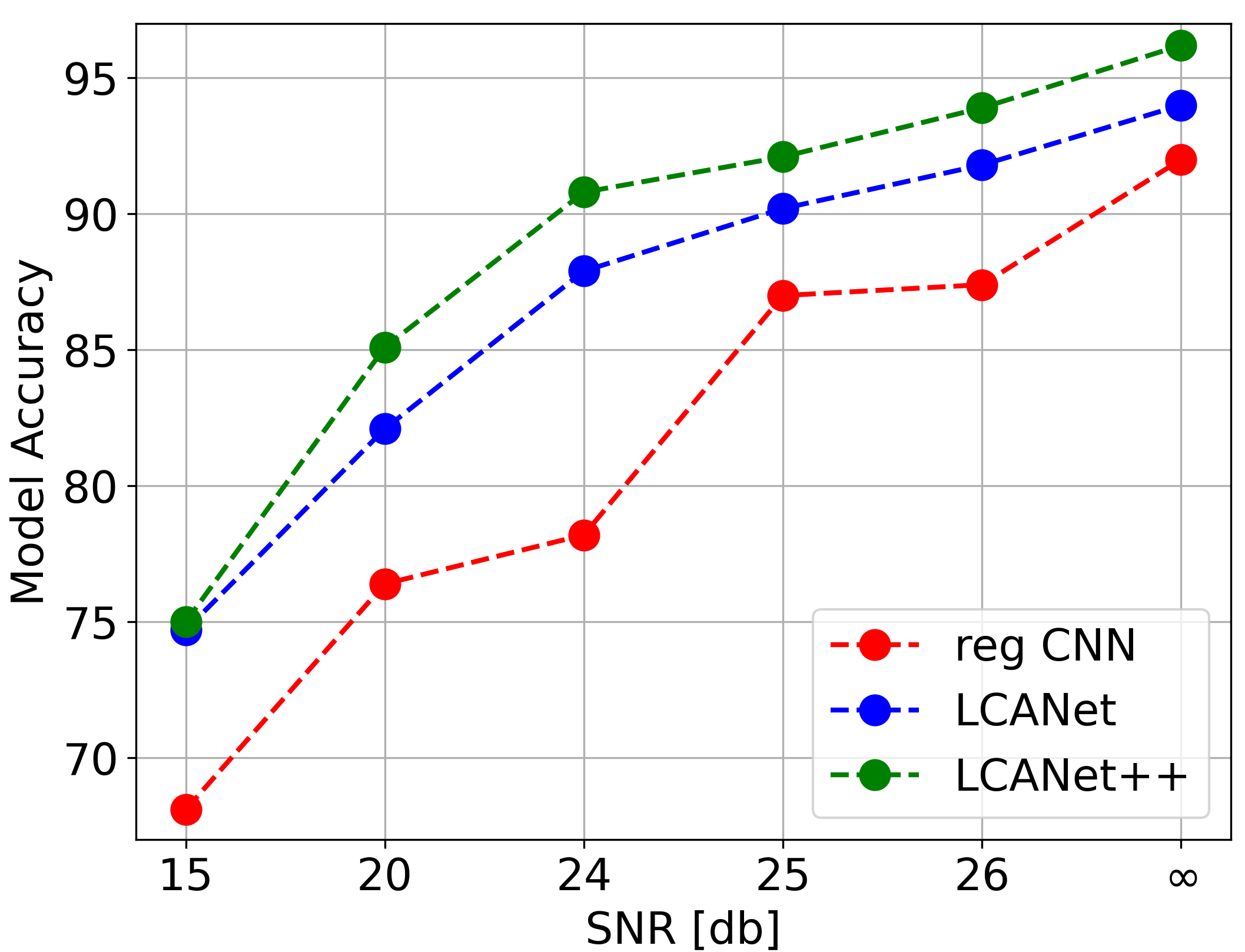}
  \caption{Regular CNN}
  \label{fig:res1}
\end{subfigure}%
\begin{subfigure}{.24\textwidth}
  \centering
  \includegraphics[width=4.0cm]{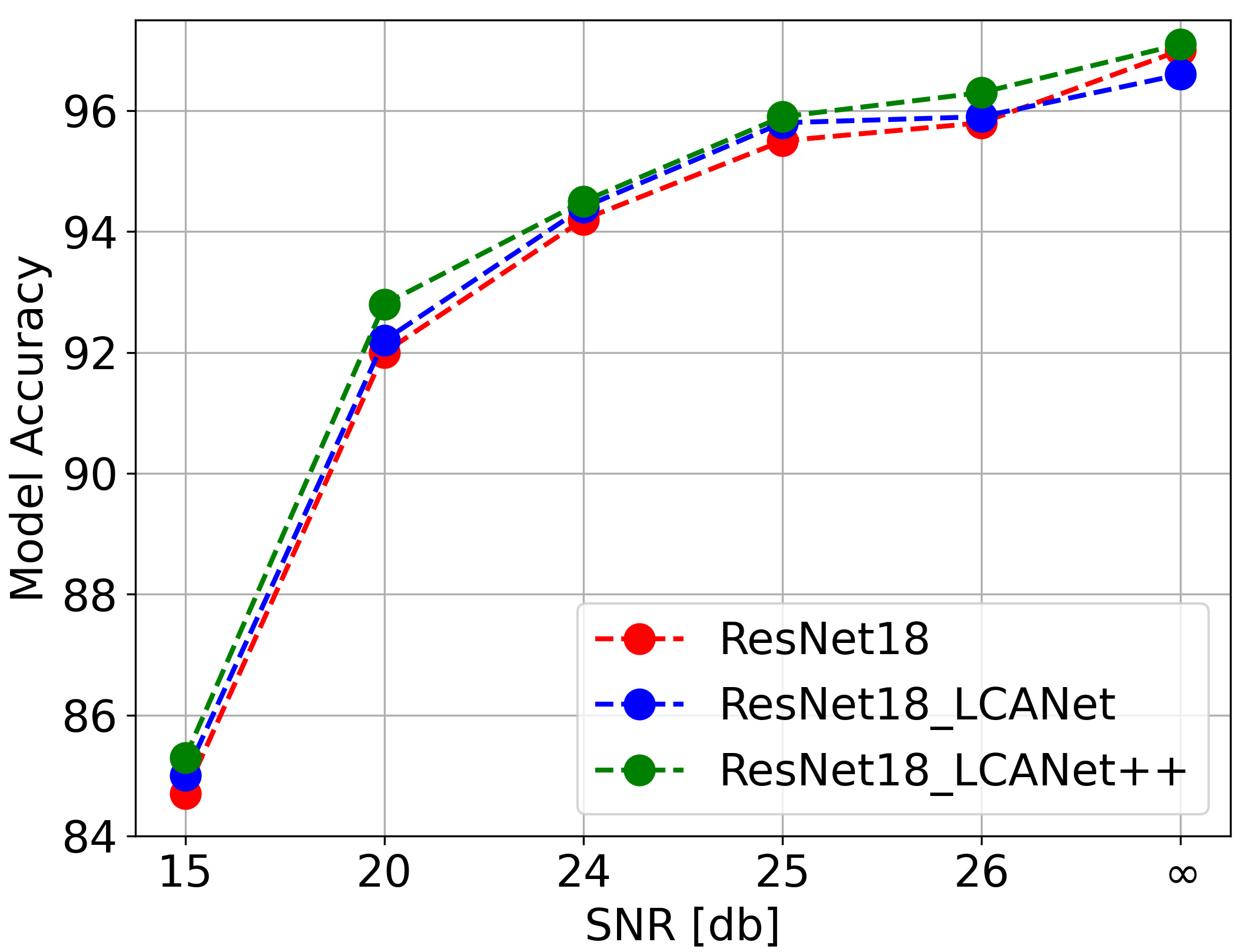}
  \caption{ResNet18}
  \label{fig:res2}
\end{subfigure}
\caption{Comparisons of our LCANets++ and other SOTA models against  perturbations with \textit{background noise}.}
\label{fig:performance_lcanets++}
\end{figure}

\subsection{Input Perturbations}
\label{subsec:adv_ptr}
We test the robustness of standard CNNs without the LCA layer(s), LCANets, and our proposed  LCANets++ to perturbations. We experiment with two different cases of input perturbations: i) \textit{background noise} on the raw audio clips and ii) \textit{gaussian noise} on MFCCs to compare robustness against both perturbation scenarios. 

\subsubsection{Background Noise}
\label{subsubsec:ptr_bn}
In Fig.~\ref{fig:res1}, we present the performance of regular CNN, LCANet, and our proposed LCANet++, tested on different perturbed test sets with \textit{background noise}, varying SNR [db] values. Observe that the regular CNN model performance drastically goes down as more perturbation is applied to original waveforms (i.e., lower SNRs). Whereas, LCANet goes down slowly with increasing perturbations, and our proposed LCANet++ shows the most robustness compared to LCANet and regular CNN models, as presented in Fig.~\ref{fig:res1}. This is attributed to the fact that LCA layers learn in an unsupervised fashion, reducing the numbers of the neurons activated through lateral competitions. These fewer activated neurons represent the most relevant input features, which are less impacted by slight perturbations. 

\begin{table}[]
\caption{Performance comparisons against  perturbations with Background Noise on waveforms}
\label{table:peformance_bn}
\centering

\begin{tabular}{  p{2.5cm} p{1.0cm} p{.6cm} p{.6cm} p{.6cm} p{.6cm}  }

\hline
 Model  & $SNR$   & &  &  & \\
 & $=15 db$   &$20 db$  &$24 db$  &$25 db$  & $\infty$\\
\hline
CNN   &0.692	&0.788	&0.793	&0.858 &0.920\\ 
 LCANet  &0.760	&0.840	&0.876	&0.904 &0.940\\ 
\textbf{LCANet++} & \textbf{0.768}  & \textbf{0.847} & \textbf{0.903} & \textbf{0.914} & \textbf{0.962}\\ 
\hline
\hline

ResNet18  & 0.847	&0.920	&0.942	&0.955 & 0.970\\ 
 ResNet18\_LCA &0.850	&0.922	&0.944	&0.958 & 0.966\\ 
\textbf{ResNet18\_LCA++} & \textbf{0.853}  & \textbf{0.928} & \textbf{0.945} & \textbf{0.959} & \textbf{0.971}\\ 

\hline

\end{tabular}
\end{table}

We also test the robustness of LCANets++ on larger models, i.e., ResNet18 model with 18 layers. As presented in Fig.~\ref{fig:res2}, we observe that the ResNet18 with multilayer LCAs, i.e., ResNet18\_LCANet++ outperforms regular ResNet18 and ResNet18 with LCA in the first layer, i.e., ResNet18\_LCANet. From Table~\ref{table:peformance_bn}, we find that for the ResNet18 architecture, LCANets++ slightly improves the robustness on perturbed test sets than regular ResNet18 without LCA layers, as opposed to significantly higher robustness LCANets++ exhibited on regular CNN model. Larger model with more layers and parameters make ResNet18  inherently more robust than regular CNNs, resulting in LCANets++ to boost up only slightly in ResNet18 than regular CNNs.

\begin{table}[]
\caption{Performance comparisons against perturbations with Gaussian Noise on MFCCs}
\label{table:peformance_gn}
\centering

\begin{tabular}{ p{1.4cm} p{.6cm} p{.6cm} p{.6cm} p{.6cm} p{.6cm} p{.6cm}  }

\hline
Model & $\epsilon$   &  &  & &  & \\
 & $=0$   & $0.01$ & $0.02$ & $0.03$ & $0.04$ & $0.05$\\
\hline
CNN   &0.866	&0.864	&0.863	&0.863	&0.858	&0.856\\ 
 LCANet    &0.939	&0.938	&0.935	&0.925	&0.909	&0.883\\ 
\textbf{LCANet++} & \textbf{0.950}  & \textbf{0.943} & \textbf{0.939} & \textbf{0.927} & \textbf{0.914} & \textbf{0.900}\\ 
\hline

\end{tabular}
\end{table}

\subsubsection{Gaussian Noise}
\label{subsubsec:ptr_gn}
We impose \textit{Gaussian noise} on the MFCCs varying $\epsilon$ values. As presented in Table.~\ref{table:peformance_gn}, with increasing the $\epsilon$ (more perturbations), performance of the regular CNN model goes down. Also, LCANet and LCANet++ performance slightly goes down, but still, the models with LCA layers show more robustness compared to the model without LCA layers, i.e., regular CNN model. This shows that our LCANets++ are more robust not only against perturbations on raw waveforms, but also against perturbations on the feature space, i.e., MFCCs.

\begin{table}[]
\caption{Comparisons against \textit{white-box} attacks}
\label{table:peformance_wbox}
\centering

\begin{tabular}{ p{.5cm} c p{.6cm} p{.6cm} p{.6cm} p{.6cm} p{.6cm}  }

\hline
Attack & Model & $\epsilon$   & &  &  & \\
 &  & $=0$   & $0.01$ & $0.016$ & $0.02$ & $0.03$\\
\hline
\multirow{3}{*}{FGSM} & CNN  & 0.866 & 0.439 & 0.196 & 0.108 & 0.017\\ 
 &LCANet &0.939   & 0.261 & 0.123 & 0.092 & 0.062\\ 
&\textbf{LCANet++} & \textbf{0.950}  & \textbf{0.679} & \textbf{0.418} & \textbf{0.417} & \textbf{0.414}\\ 
\hline
\hline
\multirow{3}{*}{PGD} & CNN  & 0.866 & 0.382 & 0.147 & 0.073 & 0.025\\ 
 &LCANet & 0.939  & 0.028 & 0.005 & 0.005 & 0.005\\ 
&\textbf{LCANet++} & \textbf{0.950}  & \textbf{0.588} & \textbf{0.585} & \textbf{0.579} & \textbf{0.567}\\ 
\hline

\end{tabular}
\end{table}

\subsection{Adversarial Attacks}
\label{subsec:adv_attacks}
We experiment with adversarial attacks having different capabilities. For experimental purposes, we consider both the \textit{white-box} and \textit{black-box} adversarial attacks.

\subsubsection{White-box Attacks}
\label{subsubsec:white_attack}
In \textit{white-box} attacks, an adversary has more capabilities like having access to the model architectures, including model parameters, weights, and gradients. We consider two different types of \textit{white-box} attacks, i.e., \textit{FGSM}~\cite{kurakin2018adversarial} and \textit{PGD}~\cite{chiang2020witchcraft}. In both attacks, the adversary utilizes the gradients to perturb the MFCCs of test sets to misclassify them during inference.  

We present the performances of the regular CNN model and LCANets, as well as our proposed LCANets++, against the FGSM attack in Fig.~\ref{fig:FGSA_attack}. We find that the regular CNN is not very robust, and it's performance goes down, as perturbations ($\epsilon$) go higher. We observe that, the single-layer LCA, i.e., LCANets are not robust against the \textit{white-box} FGSM attack, which is consistent to findings in ~\cite{teti2022lcanets} for CV tasks. However, our proposed multi-layer LCANets++ outperforms the CNN model and LCANets on audio classification agianst the FGSM attack. In Fig.~\ref{fig:FGSA_attack}, we observe that LCANets++ performance decreases comparatively slowly as attack becomes stronger with higher perturbations ($\epsilon$). We also experiment with another \textit{white-box} attack, i.e., PGD attack, where LCANets++ consistently show more robustness than SOTA models and LCANets, as shown in Table~\ref{table:peformance_wbox}.

\subsubsection{Black-box Attacks}
\label{subsubsec:black_attack}
\begin{figure}[]
%\centering
\begin{subfigure}{.24\textwidth}
  \centering
  \includegraphics[width=4.0cm]{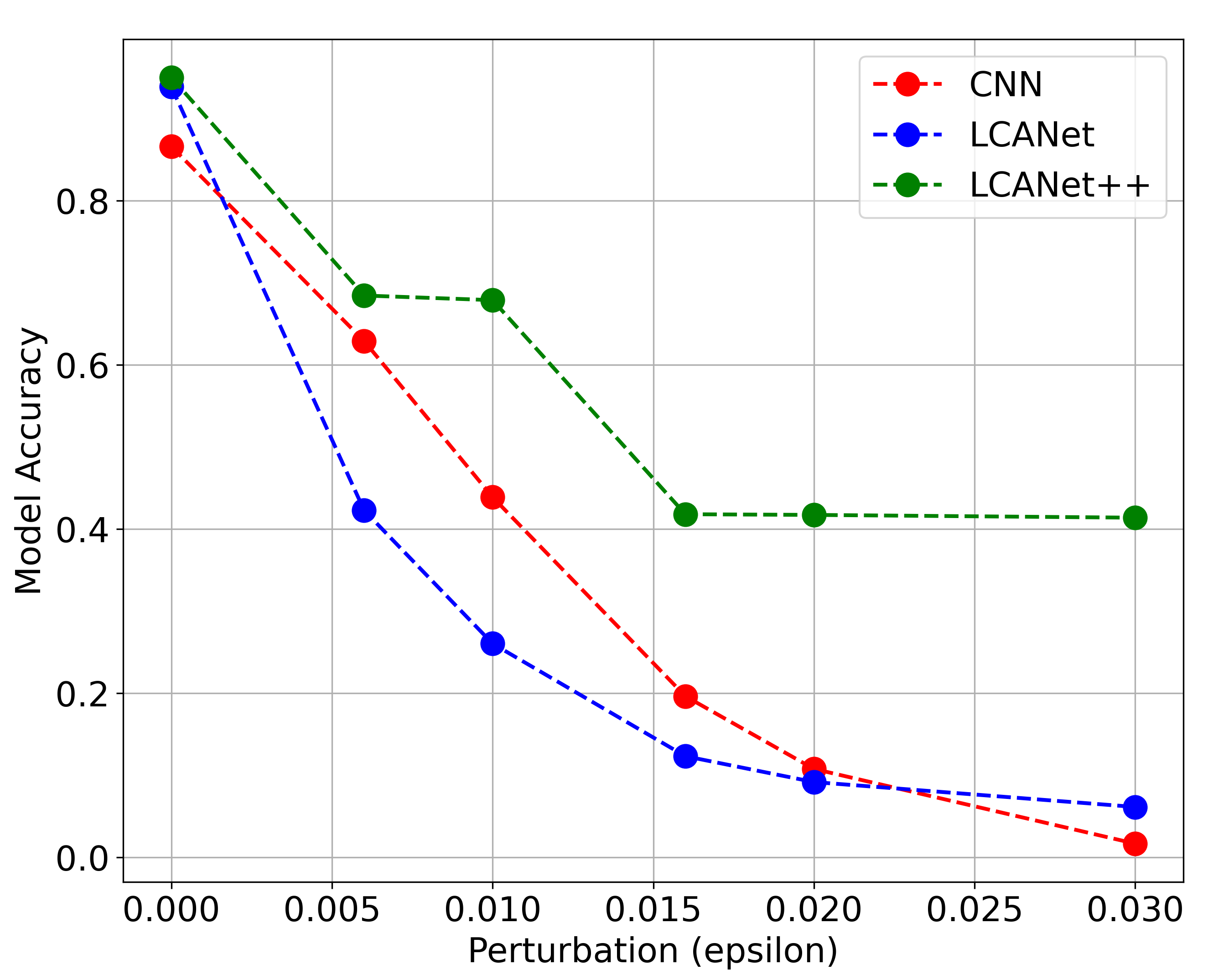}
  \caption{FGSM (\textit{white-box}) Attack}
  \label{fig:FGSA_attack}
\end{subfigure}%
\begin{subfigure}{.24\textwidth}
  \centering
  \includegraphics[width=4.0cm]{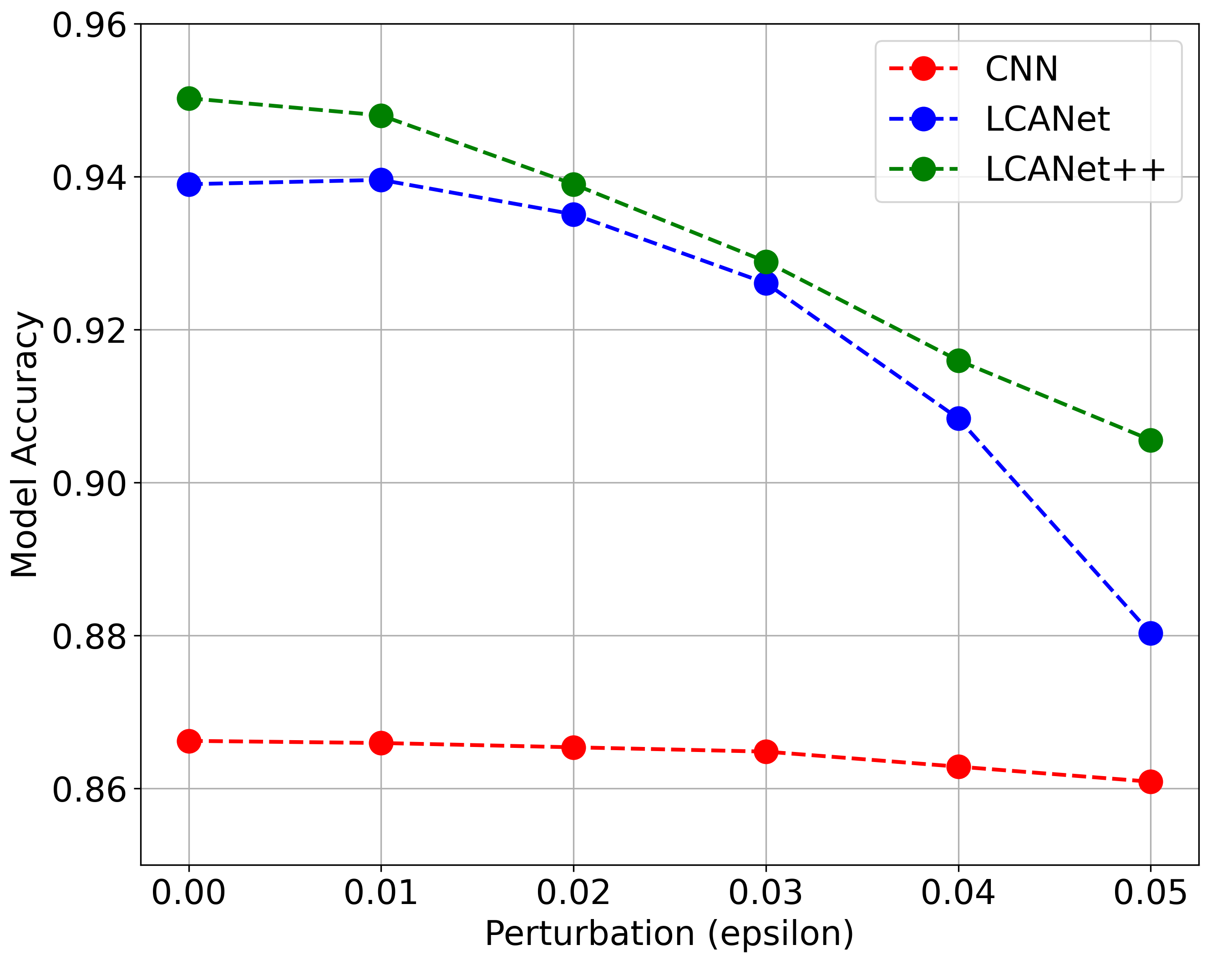}
  \caption{Evasion (\textit{black-box}) Attack}
  \label{fig:Evasion_attack}
\end{subfigure}
\caption{Comparisons of LCANets++ and SOTA models on $L_\infty$ norm \textit{white-box} attacks.}
\label{fig:White-box_attack}
\end{figure}

We experiment with the \textit{black-box} evasion attack, where the adversary has no access to the model gradients. In this attack, an adversary only has query access to the model and can get predictions from the model utilizing the query access. In our setup, the adversary is able to make queries to the original target model and get predictions from the model. The adversary utilizes the predictions and input queries to develop a surrogate model. The surrogate model generates the perturbed samples, varying perturbations ($\epsilon$), and we tested the performance of the original models on these perturbed test sets. We present the performances of regular CNN, LCANet, and our proposed LCANet++ against the \textit{black-box} evasion attack in Fig.~\ref{fig:Evasion_attack}. We observe that, in \textit{black-box} evasion attack, LCANet shows more robustness compared to CNN and LCANet++ outperforms all the models on perturbed test sets (i.e.,  $\epsilon>0$). Note that, models are trained for 20 epochs with three audio classes (i.e., limited samples), which might lead to a significant performance gap among the regular CNN and LCA-based models on unperturbed test sets  $\epsilon=0$ in Table.~\ref{table:peformance_bbox}.

\ignore{=======================

\begin{figure}
%\centering
%\begin{subfigure}{.25\textwidth}
  \centering
  \includegraphics[width=4.0cm]{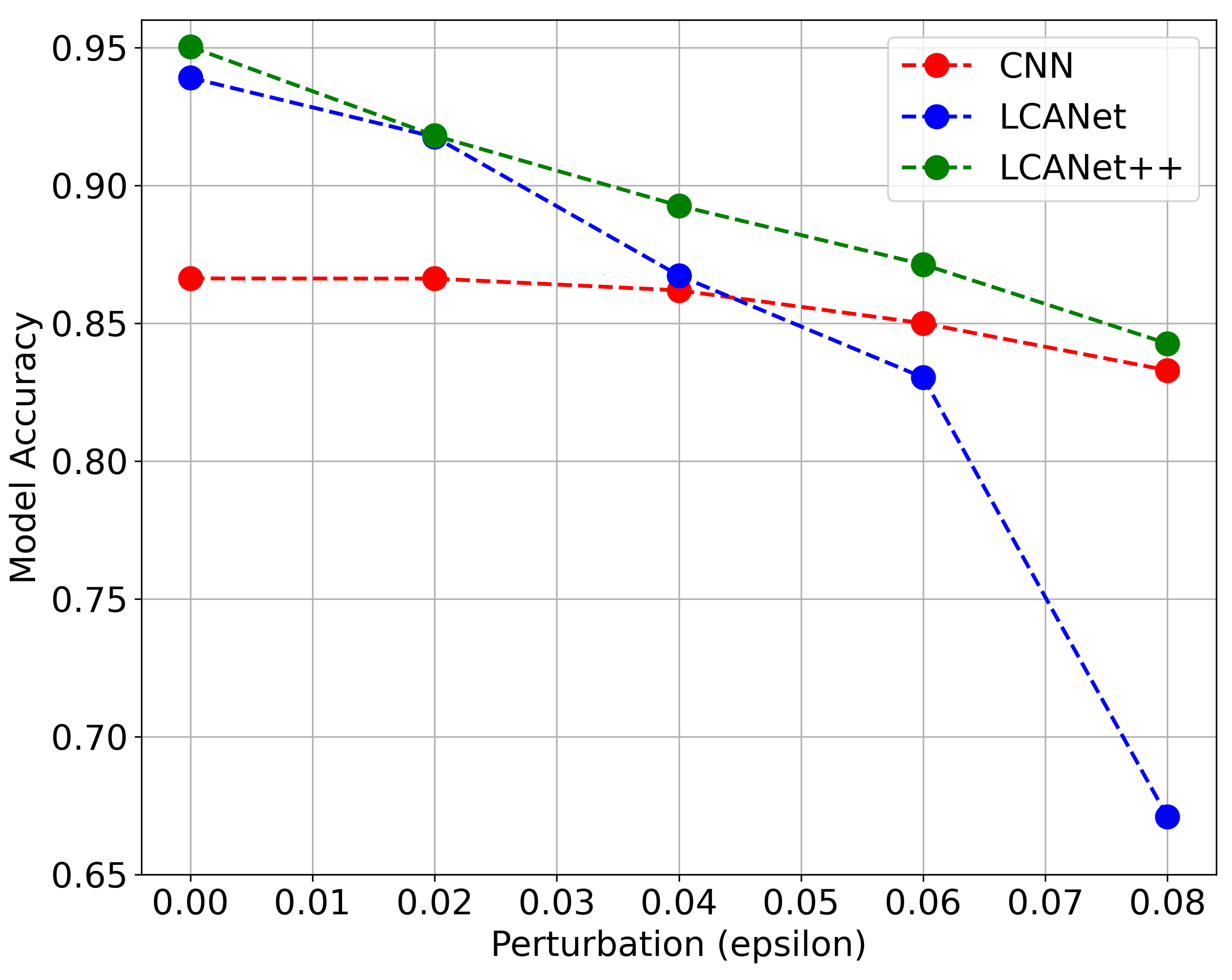}
  \caption{Evasion (\textit{black-box}) Attack}
  \label{fig:ev_attack}
\end{figure}

==================================}

\begin{table}[]
\caption{Comparisons against   \textit{black-box} (Evasion) attack}
\label{table:peformance_bbox}
\centering

\begin{tabular}{ p{1.4cm} p{.6cm} p{.6cm} p{.6cm} p{.6cm} p{.6cm} p{.6cm}  }

\hline
Model & $\epsilon$   &  &  & &  & \\
 & $=0$   & $0.01$ & $0.02$ & $0.03$ & $0.04$ & $0.05$\\
\hline
CNN   &0.866	 &0.865	 &0.865	 &0.864	 &0.862	 &0.860	\\ 
 LCANet    &0.939	 &0.939	 &0.935	 &0.926	 &0.908	 &0.880\\ 
\textbf{LCANet++} & \textbf{0.950}  & \textbf{0.948} & \textbf{0.939} & \textbf{0.928} & \textbf{0.915} & \textbf{0.905}\\ 
\hline

\end{tabular}
\end{table}

%\subsection{Ablation Study}
%\label{subsec:abl_study}

\ignore{===========

\begin{figure}
%\centering
\begin{subfigure}{.25\textwidth}
  \centering
  \includegraphics[width=4.0cm]{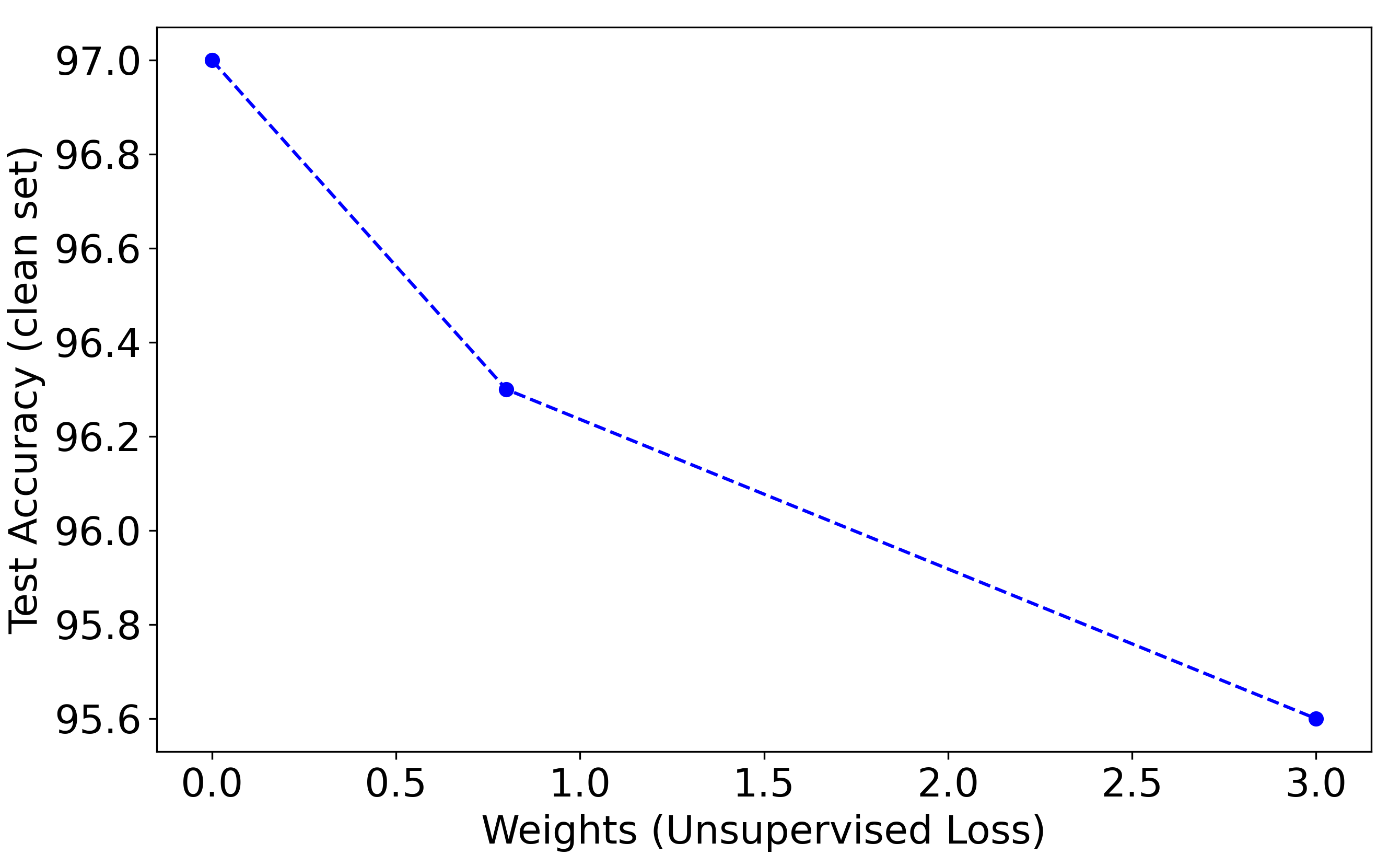}
  \caption{Clean test set}
  \label{fig:performance_clean_gn}
\end{subfigure}%
\begin{subfigure}{.25\textwidth}
  \centering
  \includegraphics[width=4.0cm]{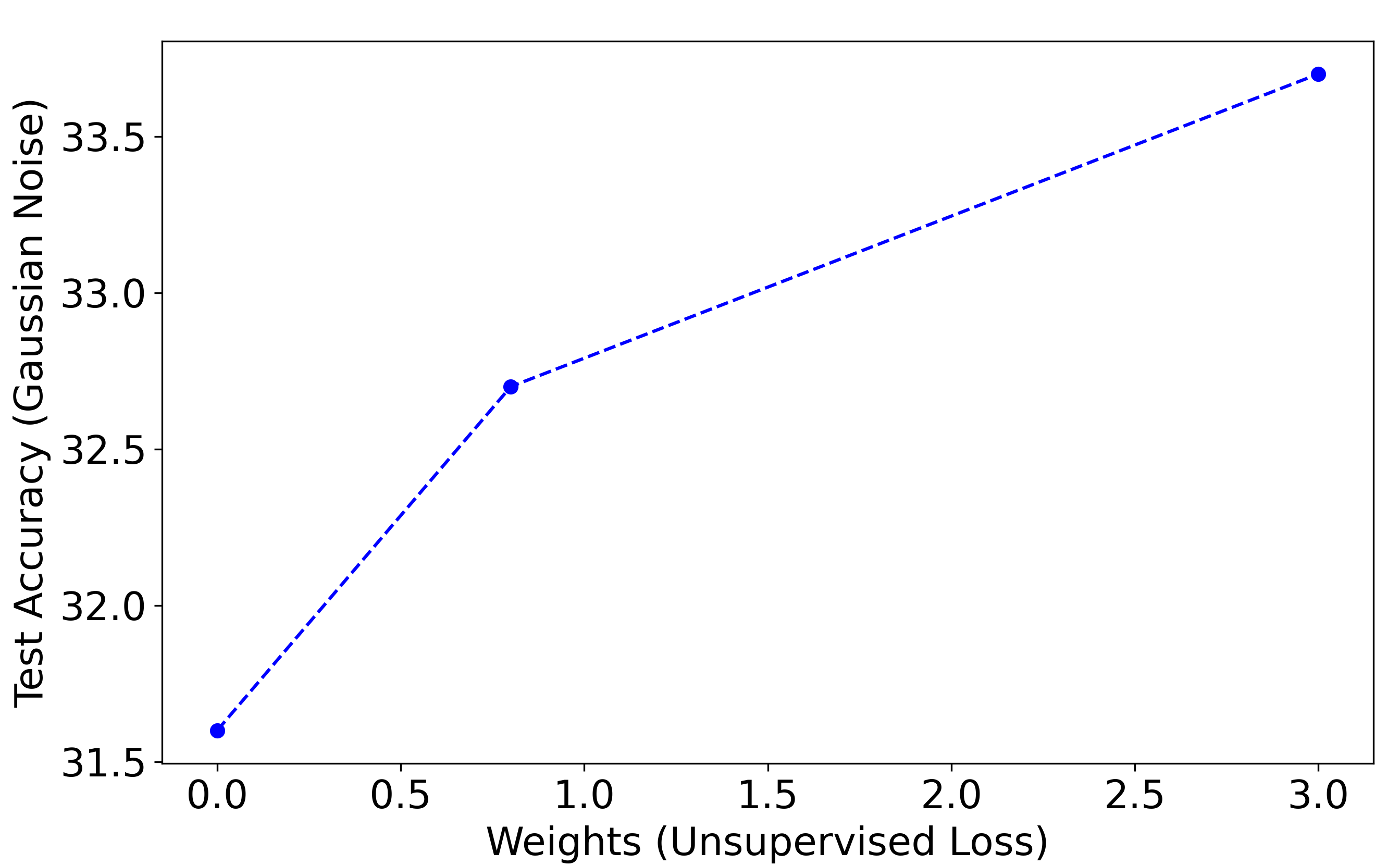}
  \caption{Perturbed test set}
  \label{fig:performance_ptr_gn}
\end{subfigure}
\caption{LCANets++ test accuracies, varying models trained with unsupervised loss weights, tested on (a.) clean test set, and (b.) Gaussian noise perturbed test sets.}
\label{fig:weight_vs_performance}
\end{figure}

======================}

%% file: Conclusion.tex
\section{Conclusions}
\label{conclusion}
In this work, we developed CNNs with sparse coding in multiple layers, referred to as LCANets++. We showed that LCANets++ can be easily implemented using regular CNNs like ResNet18. Our empirical analysis shows that LCANets++ can be used in audio classifiers to increase robustness to noise and adversarial attacks relative to LCANets and standard CNNs. In addition, we observe how the unsupervised training with LCA and number of LCA layers impacts clean and robust test accuracy. Overall, our work sheds light into future directions in designing privacy-preserving robust audio classifiers.  

%% file: Sections/6.Acknowledgements.tex
\section{Acknowledgements}

We gratefully acknowledge support from the Advanced Scientific Computing Research (ASCR) program office in the Department of Energy’s (DOE) Office of Science, award \#77902, as well as the Center for Nonlinear Studies and the Cyber Summer School at Los Alamos National Laboratory. 